\documentstyle[epsf,aaspp4,flushrt]{article}

\newcommand{\simgt}{\lower.5ex\hbox{$\; \buildrel > \over \sim \;$}}
\newcommand{\simlt}{\lower.5ex\hbox{$\; \buildrel < \over \sim \;$}}
\newcommand{\citet}[1] {\cite{#1}}
\newcommand{\citep}[1] {(\cite{#1})}
 \newcommand{\bm}[1]{\mbox{\boldmath$#1$}}
 \newcommand{\kaco}[1]{\left\langle{#1}\right\rangle}
 \newcommand{\skaco}[1]{\langle{#1}\rangle}

\newcommand{\baredth}{\;\overline{\raise1.0pt\hbox{$'$}\hskip-6pt
\partial}\;}
\newcommand{\edth}{\;\raise1.0pt\hbox{$'$}\hskip-6pt\partial\;}
\begin{document}
\title{On the Cross-Correlation between Dark Matter Distribution and
Secondary CMB Anisotropies by Ionized Intergalactic Medium}
\author{Masahiro Takada\altaffilmark{1}, and Naoshi Sugiyama}

\affil{%
Division of Theoretical Astrophysics,
National Astronomical Observatory,
2-21-1 Osawa, Mitaka, Tokyo 181-8588, Japan}
\affil{%
mtakada@th.nao.ac.jp; naoshi@th.nao.ac.jp}
\altaffiltext{1}{
Present address: 
Department of Physics and Astronomy,
University of Pennsylvania, 209 S. 33rd Street,
Philadelphia, PA 19104, USA; mtakada@hep.upenn.edu
}

\begin{abstract}
We investigate a cross-correlation between the weak gravitational
lensing field of the large-scale structure, $\kappa$, and the secondary
temperature fluctuation field, $\Delta$, of cosmic microwave background
(CMB) induced by Thomson scattering of CMB photons off the ionized
medium in the mildly nonlinear structure.  The cross-correlation is
expected to observationally unveil the biasing relation between the dark
matter and ionized medium distributions in the large-scale structure.
We develop a formalism for calculating the cross-correlation function
and its angular power spectrum based on the small angle approximation.
As a result, we find that leading contribution to the cross-correlation
comes from the secondary CMB fluctuation field induced by the quadratic
Doppler effects of the bulk velocity field $v$ of ionized medium, since
the cross-correlations with the $O(v)$ Doppler effect and the
Ostriker-Vishniac effect of $O(v\delta)$ are suppressed on relevant
angular scales because the linear dependence on the bulk velocity field
leads to cancellations between positive and negative contributions among
the scatterers for the ensemble average.  The magnitude of the
cross-correlation can be estimated as $\kaco{\kappa\Delta}\sim 10^{-10}$
on small angular scales ($l\simgt 1000$) under the currently favored
cold dark matter model of the structure formation and the simplest
scenario of homogeneous reionization after a given redshift of $z_{\rm
ion}\simgt 5$.  Although the magnitude turns out to be small, we find
several interesting aspects of the effect. One of them is that the
density-modulated quadratic Doppler effect of $O(\delta_e v^2)$ produces
the cross-correlation with the weak lensing field that has {\em linear}
dependence on the electron density fluctuation field $\delta_e$ or
equivalently on the biasing relation between the dark matter and
electron distributions. In other words, the angular power spectrum could
be both positive or negative, depending on the positive biasing or
antibiasing, respectively.  Detection of the cross-correlation thus
offers a new opportunity to observationally understand the reionization
history of intergalactic medium connected to the dark matter clustering.

\end{abstract}

\section{INTRODUCTION}

The absence of the hydrogen Gunn-Peterson trough in quasar absorption
spectra (\cite{Gunn}) strongly indicates that the smoothly distributed
hydrogen in the intergalactic medium (IGM) is already highly ionized by
$z=6$ (e.g., \cite{Lanzetta}; \cite{Fan}). Recently, Becker et
al. (2001) reported the first detection of the Gunn-Peterson test
through in a $z=6.28$ QSO spectrum. However, even a small fraction of
neutral hydrogen in the IGM would result in an undetectable flux because
of the large cross section to Ly$\alpha$ photons, and thus further
careful investigations will be necessary.  These results also imply the
presence of very luminous ionizing sources such as QSOs and high mass
stars at high redshifts to cause photoionizations of hydrogen by
$z\simgt 6$.  On the other hand, it is known that earlier reionization
would have brought the last scattering surface to lower redshift,
smoothing the intrinsic anisotropies in cosmic microwave background
(CMB) radiation generated at the recombination epoch $z\sim 1000$
(\cite{Bond91}; \cite{SugSilk}).  The recent measurements of the CMB
anisotropies on subdegree angular scales (\cite{Boom1}; \cite{Maxima};
\cite{Boom2}) can be used to set an upper limit on reionization redshift
such as $z_{\rm ion}\sim 30$ (\cite{Teg}). Thus, under the currently
favored adiabatic cold dark matter (CDM) scenario for the structure
formation, reionization of the universe is expected to occur in the
range of $6\simlt z_{\rm ion}\simlt 30$. Revealing how reionization
proceeds in the inhomogeneous universe stands out as one of the most
important problems in the observational cosmology (e.g., see
\cite{Barkana} for a recent review).

Secondary CMB anisotropies should be induced by scattering of CMB
photons off electrons in the ionized medium undergoing bulk motion in
the large-scale structure.  It is therefore expected that the secondary
anisotropies can provide a laboratory for studying structure formation
in the reionized epoch. Kaiser (1984) investigated the CMB distortion
induced by the bulk velocity to pure linear order of $O(v)$ and found
that the induced anisotropies are small at arcminite scales resulting
from cancellations between positive and negative shifts along the line
of sight, since the gravitationally induced bulk velocities are
irrotational.  Ostriker \& Vishniac (1986) then pointed out that the
coupling of the Doppler effect with spatial variations in the free
electron density produces $\mu K$ anisotropies on arcminite scales at
the order of $O(\delta v)$, which is the so-called Ostriker-Vishniac
(OV) effect, because the effect is a second-order contribution of
perturbations that does not suffer from the cancellation (see also
\cite{Vish87}; \cite{Efstathiou}; \cite{HuWhite}; \cite{Jaffe}). For the
same reason, spatial inhomogeneities of ionization fraction associated
with a realistic patchy reionization scenario should lead to the
secondary CMB anisotropies by the Doppler effect (\cite{Aghanim};
\cite{GH}; \cite{Knox}; \cite{Benson}; \cite{Gnedin01}). In particular,
Benson et al. (2001) investigated the secondary CMB fluctuations under a
realistic model of the reionization of IGM caused by starts in
high-redshift galaxies, based on a semi-analytic model of galaxy
formation and a high-resolution N-body simulation of the dark matter
clustering. They showed that measurements of CMB anisotropies could
actually put stringent constraints on the reionization processes. On the
other hand, from theoretical aspects, it is not so evident to answer the
problem which second-order effect of perturbations produces dominant
contribution to the secondary CMB fluctuations. Hu, Scott, \& Silk (1994
hereafter HSS; see also Dodelson \& Jubas 1995) derived a complete
Boltzmann equation including the quadratic Doppler effects $O(v^2)$ of
bulk velocity that governs redshift evolution of the CMB temperature
fluctuation field. Consequently, they found that, although the quadratic
Doppler effect should induce spectral distortion of CMB like the thermal
Sunyaev-Zel'dovich (SZ) effect (\cite{SZ}) as a new qualitative feature,
the magnitude is always smaller than that of the OV effect for the CDM
models.
  
A great uncertainty in models for structure formation in the reionized
epoch is the extent and nature of reionization, because this requires
the understanding of various complicated astrophysical processes such as
the nature of first ionizing sources, the fraction of ionizing photons
that escape the sources to become available for photoionization of the
hydrogen in IGM, and the radiative balance between the escaped
emissivity and the recombination rate in the inhomogeneous IGM. For
example, even a naive question which high-density or low-density region
in the large-scale structure becomes first photoionized bas not still
been answered definitely. Assuming {\it ab initio} that the distribution
of the ionization fraction is given as a function of gas density,
Miralde-Escud\'e, Hanehnelt, \& Rees (2000; hereafter MHR) suggested
that the UV radiation may escape into and ionize the low-density region,
where the recombination rates are the lowest, even though it is expected
that the ionizing sources could be associated with the peaks in the
density field. This conclusion has also been confirmed by the recent
numerical simulations taking into account the radiative transfer
approximately (\cite{Gnedin00}).

While the relation between the dark matter and ionized medium
distributions is thus a crucial clue to understanding the reionization
history connected with the gravitational instability scenario of
structure formation, the relation seems difficult to be directly
observed by methods so far proposed.  Therefore, the purpose of this
paper is to investigate the cross-correlation between the dark matter
distribution and the secondary CMB temperature fluctuation field induced
by the ionized medium in the mildly nonlinear structure.
It is now recognized that the weak gravitational lensing effect due to
the large-scale structure can be a unique tool to observationally
reconstruct the dark matter distribution (e.g., see \cite{Bartel} for a
review). There are two feasible methods often considered in the
literature; one is using the distortion effects on distant galaxy images
that have been actually detected (e.g., \cite{Ludvic}), and the other is
using the weak lensing effects on the CMB map (\cite{SZPRL};
\cite{ZS99}; \cite{TKF}; \cite{TF}; \cite{STF}; \cite{Takada01}; 
\cite{HuLens01},b).  In
particular, it is expected that the method of using the distortion
effects on the CMB map developed by Seljak \& Zaldarriaga (1999; also
\cite{ZS99}) allows us to reconstruct the projected dark matter
distribution with high accuracy. Based on these considerations, in this
paper we assume that the weak lensing field is {\it a priori}
reconstructed by those methods, for example, using the future CMB data
from the satellite mission {\em Planck Surveyor}.  Then, taking a
cross-correlation between the weak lensing field and the CMB map
including the secondary fluctuations would be a challenging test.
Furthermore, the cross-correlation method could improve the
signal-to-noise (S/N) ratio of the Doppler effect compared with the
original S/N in the CMB map itself if the secondary CMB field correlates
with the weak lensing field, which is analogous to, for example, the
cross-correlation between the CMB temperature fluctuation and
polarization fields (\cite{HuWhite}; see also \cite{Peiris} and
\cite{Ludvic00} for other examples).  In this paper, based on the small
angle approximation, we first develop a formalism for calculating the
cross-correlation function and its angular power spectrum between the
weak lensing field and the secondary CMB fluctuation field including the
Doppler effects up to the second-order of the bulk velocity field. We
then estimate the magnitude of the cross-correlation under the currently
favored CDM models and the simplest assumption of homogeneous
reionization history after some epoch $z_{\rm ion}\simgt 5$.

The layout of this paper is as follows. In \S \ref{Prel} we review the
relevant properties and parameters of the adiabatic CDM scenario for
structure formation. In \S \ref{Boltz} we briefly review the Boltzmann
equation including the Doppler effect up to the second-order of bulk
velocity and discuss the physical implications of the effect following
HSS.  In \S \ref{Form} we develop a formalism for calculating the
cross-correlation function and its angular power spectrum between the
weak lensing field and the secondary CMB temperature fluctuation field
induced by the bulk motion of the ionized medium in the large-scale
structure.  Results for the currently favored CDM models are presented
in \S \ref{result}. In \S \ref{Conc} we present the discussions and
conclusions. Throughout this paper, we employ unit of $c=1$ for light
speed $c$.

\section{PRELIMINARIES}
\label{Prel}
\subsection{FRW Model}
The expansion rate of the universe can be described by the Friedmann
equation:
\begin{equation}
H(t)\equiv\frac{1}{a}\frac{da}{dt}=H_0\sqrt{\Omega_{\rm m0}a^{-3}
+\Omega_{\lambda0}},
\end{equation}
where $a(t)$ denotes the scale factor, $H_0=100h{~\rm km~s}^{-1}{\rm
Mpc}^{-1}$ is the Hubble constant, $\Omega_{\rm m0}$ and
$\Omega_{\lambda0}$ denote the density parameters of the present-day
non-relativistic matter and the contribution of the cosmological
constant relative to the critical density $\rho_{\rm cr0}=3H_0^2/8\pi
G$, respectively. Here we have set $a(t_0)=1$ at present for
convenience.  Throughout this paper, we consider only a flat universe
model as supported by the recent high precision CMB experiments (e.g.,
\cite{Boom1}; \cite{Boom2}) and then we have
$\Omega_{\lambda0}=1-\Omega_{\rm m0}$. The conformal time which we will
often use in the following discussions is defined by $d\eta=a^{-1}dt$.
It is convenient to introduce the comoving angular diameter distance
$r(\eta)$ with respect to $\eta$ or equivalently redshift $z$,
and for a flat universe model it is given by
\begin{equation}
r(\eta)=\eta_0-\eta=\int^{z}_0\!\!\frac{dz}{H(z)},
\end{equation}
where $\eta_0$ is the present conformal time and $z$ is given by 
$1+z=1/a$.

It is currently believed that the reionization of the universe occurred
at $6\simlt z_{\rm ion}\simlt 30$ such that the reionized medium is
optically thin to the Thomson scattering of CMB photons, which is
characterized as $\tau< 1$ in terms of the optical depth, $\tau$.
Assuming homogeneous ionization after some reionization epoch $z_{\rm
ion}$, $\tau$ can be analytically expressed as a function of $z$ for 
a flat universe  model:
\begin{equation}
\tau\equiv\int^{t_0}_t\!\!dt' \bar{n}_e\sigma_T 
=\frac{2}{3}\frac{\tau_{\rm H}\bar{X}_e}{\Omega_{\rm m0}}\left[
\sqrt{1-\Omega_{\rm m0}+\Omega_{\rm m0}(1+z)^3}-1\right], 
\label{eqn:opt}
\end{equation}
with 
\begin{equation}
\tau_{\rm H}\equiv \bar{n}_{e0}\sigma_{\rm T}H_0^{-1} 
=0.0692(1-Y)\Omega_{\rm b0}h.
\end{equation}
Here $\sigma_{\rm T}$ is the Thomson cross section, $\bar{n}_e$ is the
mean number density of free electron, $\bar{X}_e$ is the mean ionization
fraction, which does not depend on time here, $\Omega_{\rm b0}$ is the
present-day density parameter of baryon in units of the critical
density, $\tau_H$ is the optical depth to Thomson scattering to the
Hubble distance today under the assumption of full hydrogen ionization
and homogeneous distribution of baryon, and $Y$ is the primordial helium
fraction. We fix $Y=0.24$ throughout this paper.  

Again in the case of homogeneous and full hydrogen ionization, the 
visibility function that gives the probability of last scattering within
$d\eta$ of $\eta$ is expressed as
\begin{equation}
g(\eta)=\dot{\tau}e^{-\tau}=\bar{n}_e \sigma_{\rm T}a e^{-\tau}= 
\tau_{\rm H} H_0a^{-2}e^{-\tau},
\label{eqn:visib}
\end{equation}
where overdot denotes derivative with respect to the conformal time
$\eta$. 
  
\subsection{Perturbation Theory of the Large-Scale Structure Formation}
\label{LSS} The evolution of the large-scale structure after the
decoupling epoch $z_\ast\sim 1000$ is characterized by the time
evolution and scale dependence of the density contrast field
$\delta(\bm{x},t)$ mainly caused by the dark matter distribution.  The
gravitational instability then leads to bulk motions of structure with
the velocity field $\bm{v}(\bm{x})$, and we expect that the ionized
baryonic medium also follows the bulk motion on large scales, leading to
the Doppler effects on CMB photons through the Compton scattering by
electrons. Since we focus our investigations on the secondary CMB
fluctuations up to the second-order of $v$, we need to consider the
fields of $\delta$ and $\bm{v}$ up to the second-order that are valid in
the mildly nonlinear regime such as $\delta\simlt 1$:
\begin{eqnarray}
&&\delta(\bm{x})=\delta^{(1)}(\bm{x})+\delta^{(2)}(\bm{x})+\cdots,\nonumber\\
&&\bm{v}(\bm{x})=\bm{v}^{(1)}(\bm{x})+\bm{v}^{(2)}(\bm{x})+\cdots.
\label{eqn:pert}
\end{eqnarray}
The quantities with superscripts `(1)' and `(2)' denote the first- and
second-order perturbation quantities, respectively.  The time
dependences of these fields can be expressed in terms of cosmological
parameters based on the (Eulerian) perturbation theory (e.g.,
\cite{Peebles}).  In the linear regime, the time dependence of
$\delta^{(1)}$ is given by the growing factor as
$\delta^{(1)}(\bm{x},t)= D(t)\delta^{(1)}(\bm{x})$ with 
\begin{equation}
D(t)\propto H(t)\int^{1}_{a(t)}\!\!\frac{da}{(Ha)^3},
\end{equation}
where we have neglected the radiation energy contribution, and will
normalize $D(t_0)=1$ at the present time $t_0$ for convenience in the
followings.

The Fourier transformation of $\delta^{(1)}(\bm{x})$ is expressed as
\begin{equation}
\delta^{(1)}(\bm{x})=\int\!\!\frac{d^3\bm{k}}{(2\pi)^3}\delta^{(1)}_{\bm{k}}
e^{i\bm{k}\cdot \bm{x}}. 
\end{equation}
If the linear density fluctuations are Gaussian, the statistical
properties can be completely specified by the power spectrum $P(k)$
defined by
\begin{equation}
\skaco{\delta_{\bm{k}}\delta^\ast_{\bm{k}'}}\equiv (2\pi)^3
P(k)\delta^3(\bm{k}-\bm{k}'),
\end{equation}
where $\kaco{\dots}$ denotes an ensemble average over all realizations.
Note that time evolution of the power spectrum is given by
$P(k,t)=D^{2}(t)P(k)$.  In this paper, assuming the adiabatic CDM
scenario, we employ the Harrison-Zel'dovich spectrum and the BBKS
transfer function (\cite{BBKS}) with the shape parameter given by
Sugiyama (1995) as for the shape of $P(k)$. The free parameter is then
the normalization of $P(k)$ only, which is conventionally expressed in
terms of the rms mass fluctuations of a sphere of $8h^{-1}$Mpc radius,
i.e., $\sigma_8$.  From the continuity equation, we can obtain the
Fourier components of the bulk velocity filed $\bm{v}^{(1)}_{\bm{k}}$ at
the first order as
\begin{equation}
\bm{v}^{(1)}_{\bm{k}}(t)=\frac{i\bm{k}}{k^2}\dot{D}\delta^{(1)}_{\bm{k}}. 
\label{eqn:pecvel}
\end{equation}

Likewise, based on the perturbation theory, we can obtain the 
second-order Fourier components of density fluctuation and bulk velocity fields 
in terms of $\delta^{(1)}_{\bm{k}}$ and $D$ (e.g., see \cite{Peebles}):
\begin{eqnarray}
&&\delta^{(2)}_{\bm{k}}(t)=D^2\int\!\!\frac{d^3\bm{k}'}{(2\pi)^3}
F(\bm{k}',\bm{k}-\bm{k}')\delta^{(1)}_{\bm{k}'}\delta^{(1)}_{\bm{k}-\bm{k}'}, 
\nonumber\\
&&\bm{v}^{(2)}_{\bm{k}}(t)=\frac{i \bm{k}}{k^2}D\dot{D}
\int\!\!\frac{d^3\bm{k}'}{(2\pi)^3}G(\bm{k}',\bm{k}-\bm{k}')
\delta^{(1)}_{\bm{k}'}\delta^{(1)}_{\bm{k}-\bm{k}'},
\label{eqn:2pert}
\end{eqnarray}
with 
\begin{eqnarray}
&&F(\bm{k}_1,\bm{k}_2)\equiv\frac{5}{7}+\frac{1}{2}\left(\frac{k_1}{k_2}
+\frac{k_2}{k_1}\right)\frac{\bm{k}_1\cdot\bm{k}_2}{k_1k_2}+
\frac{2}{7}\left(\frac{\bm{k}_1\cdot\bm{k}_2}{k_1k_2}\right)^2,
\nonumber\\
&&G(\bm{k}_1,\bm{k}_2)\equiv\frac{3}{7}+\frac{1}{2}\left(\frac{k_1}{k_2}
+\frac{k_2}{k_1}\right)\frac{\bm{k}_1\cdot\bm{k}_2}{k_1k_2}+
\frac{4}{7}\left(\frac{\bm{k}_1\cdot\bm{k}_2}{k_1k_2}\right)^2,
\end{eqnarray}
where we have neglected extremely weak dependences of the functions $F$
and $G$ on cosmological parameters $\Omega_{\rm m0}$ and
$\Omega_{\lambda0}$ for simplicity, and considered the scalar type
perturbations only.  It is worth noting that, even if perturbations
start from Gaussian in the linear regime, the mildly nonlinear evolution
leads to non-vanishing third-order moments such as $\skaco{\delta^3}$
and $\skaco{\delta v^2}$, because those quantities are essentially the
fourth-order moments of linear perturbation quantities like
$\skaco{(\delta^{(1)})^4}$.

\section{BOLTZMANN EQUATION FOR SECOND-ORDER COMPTON SCATTERING}
\label{Boltz} Here, following HSS, we briefly review the derivation of
the secondary CMB temperature fluctuation field induced by the Doppler
effect of ionized medium up to the second order $O(v^2)$ of the bulk
velocity.  Let us first write down the Boltzmann equation in terms of
the CMB photon energy $p$ and direction $\gamma^i$ in the presence of
Compton scattering between the CMB photons and free electrons that could
lead to energy changes;
\begin{equation}
\frac{\partial f}{\partial t}+\frac{\partial f}{\partial x^i}
\frac{d x^i}{dt}+\frac{\partial f}{\partial p}\frac{dp}{dt}+
\frac{\partial f}{\partial \gamma^i}{d \gamma^i}{dt}=C(x,p),
\label{eqn:Boltz1}
\end{equation}
where $f$ is the photon occupation number and $C(x,p)$ is the collision
term by the Compton scattering. HSS derived the collision term that
leads to the Doppler effect up to the second order $O(v^2)$ as 
\begin{equation}
C_{v}(x,p)=n_{\rm e}\sigma_{\rm T}\left[-\bm{\gamma}
\cdot\bm{v}p\frac{\partial f}{\partial p}+
\left\{[(\bm{\gamma}\cdot\bm{v})^2+v^2]p\frac{\partial f}{\partial p}
+\left(\frac{3}{20}v^2+\frac{11}{20}(\bm{\gamma}\cdot\bm{v})^2\right)
p^2\frac{\partial^2f}{\partial p^2}\right\} \right],
\label{eqn:collision}
\end{equation}
where we have ignored the term by thermal electron scattering, leading
to the thermal SZ effect. The first term in the large brackets on the
r.h.s represents the Doppler effect of $O(v)$ that produces the OV
effect from the coupling with the electron density fluctuation field
$\delta_e$ at the order $O(v\delta_e)$.  The other terms give the
quadratic Doppler effect of $O(v^2)$. It is interesting that the effect
leads to energy exchanges between CMB photons and electrons like the
thermal SZ effect, and therefore causes the Compton $y$-distortion. As
shown by HSS, we can get more physical insight into this effect by
considering the limit of many scattering blobs or equivalently the
multiple scattering limit.  If we can average equation
(\ref{eqn:collision}) over the direction of bulk velocity field
$\bm{v}$, the $O(v)$ Doppler effect primarily cancels in this case and
then we can obtain
\begin{equation}
C_v(x,p)=n_{\rm e}\sigma_{\rm T}\frac{\skaco{v^2}}{3}\frac{1}{p^2}
\frac{\partial}{\partial p}\left[p^4\frac{\partial f}{\partial p}
\right]. 
\label{eqn:manyblob}
\end{equation}
If introducing the effective temperature $T_{\rm eff}$ defined by
$\skaco{v^2}=3T_{\rm eff}/m_{\rm e}$ ($m_e$ is the electron mass), this
collision term reduces to the same form of that in the Kompaneets
equation with the thermal temperature $T_e$ of hot electron, which gives
the thermal SZ effect.  This implies that many ionized blobs distributed
in the large-scale structure can be considered like hot electrons inside
one cluster, and thus those blobs induce the energy exchange to the CMB
photons characterized by the effective temperature $T_{\rm eff}$.  Under
the CDM scenario for the structure formation, the amplitude of Compton
$y$-distortion induced by the quadratic Doppler effect is below that
due to the thermal SZ effect mainly caused by intracluster hot plasma.
This is because we can typically estimate $T_{\rm eff}\sim 10^{-1}{\rm
eV}$ for the gravitationally induced bulk flow and $T_e \sim {\rm keV}$
for intracluster hot plasma, and the fact of $T_e\gg T_{\rm eff}$ leads
to the conclusion, even though clusters are rare objects in the
universe.

In this paper, we do not employ the many scattering blobs limit for the
purpose of generality as done by HSS. Namely, we shall no longer be
concerned with frequency dependence of the temperature fluctuation
field. Integration of equation (\ref{eqn:Boltz1}) over photon energy $p$
leads to the Boltzmann equation up to the order $O(v^2)$ to govern the
time evolution of temperature fluctuation field, $\Delta(\equiv \delta
T/T_{\rm CMB})$, along the line of sight with direction of
$\bm{\gamma}$:
\begin{equation}
\dot{\Delta}+\gamma^i\partial_i\Delta=n_{\rm e}\sigma_{\rm T}a
\left[\Delta_0-\Delta
+\bm{\gamma}\cdot\bm{v}-v^2+7(\bm{\gamma}\cdot\bm{v})^2
\right],
\end{equation}
where we have neglected the gravitational effect and the correction due
to the quadrupole temperature fluctuations.  By solving the equation
above formally, we can obtain the secondary fluctuation field induced by
the Doppler effect of ionized medium during propagations of the CMB
photons from the decoupling $z_\ast\sim 1000$ to present:
\begin{equation}
\Delta(\eta_0,\bm{\gamma})=\int^{\eta_0}_{\eta_\ast}\!\!d\eta
g(\eta)X_{e}(\eta)[1+\delta_e(\bm{x},\eta)]
\left[\bm{\gamma}\cdot\bm{v}-v^2+7(\bm{\gamma}\cdot\bm{v})^2
\right],
\label{eqn:fullCMB2nd}
\end{equation}
where $\eta_\ast$ is the conformal time at the decoupling epoch
$z_\ast\approx 1000$, $g(\eta)$ is the visibility function given by
equation (\ref{eqn:visib}), and we have used the fact that the {\em
free} electron density field can be expressed as
\begin{eqnarray}
n_e(\bm{x},\eta)&=&\bar{n}_e(\eta)[1+\delta_e(\bm{x},\eta)]\nonumber\\
&=&\bar{n}_{\rm H0}a^{-3}(\eta)X_e(\eta)(1+\delta_e).
\label{eqn:elenum}
\end{eqnarray}
The $X_e(\eta)$ is the ionization fraction as a function of $\eta$,
$\delta_e$ is the density fluctuation field of electron along the line
of sight, and $\bar{n}_{\rm H0}$ is the mean hydrogen number density at
present given by $\bar{n}_{\rm H0}=1.12\times 10^{-5}{\rm cm}^{-3}\Omega_{\rm
b0}h^2(1-Y)$. Note that, as explicitly introduced by Gnedin \& Jaffe
(2001), $n_e$ can also be expressed as $n_e=X_e(1+\delta_X)n_{\rm
H}=X_e\bar{n}_{\rm H}[1+\delta_X(\bm{x},\eta)]
[1+\delta_b(\bm{x},\eta)]$, where $\delta_X$ and $\delta_b$ are the
fluctuation fields of the ionization fraction and the baryon density,
respectively, as a function of the spatial position and time. It is thus
clear that $\delta_e$ is different from $\delta_b$ in a general
reionization history.
Note that we will deal with $\delta_e$ throughout this paper.

\section{FORMALISM: CROSS-CORRELATION BETWEEN DOPPLER EFFECTS ON CMB AND
 WEAK LENSING FIELD}
\label{Form}

In this section, we develop a formalism for calculating the
cross-correlation function and its angular power spectrum between 
the weak lensing field and the secondary CMB anisotropies field induced
by the Doppler effect of the ionized medium.
 \subsection{Weak Lensing Field}
The weak gravitational lensing due to the large-scale structure is now
recognized as a unique tool for directly measuring the dark matter
distribution. Seljak \& Zaldarriaga (1999) (also see \cite{ZS99})
developed a useful method for reconstructing the projected field of dark
matter fluctuations between the last scattering surface and present from
the lensing distortions to the product of gradients of the temperature
filed. This method could be successfully performed by the satellite
mission Planck.  It has been also shown that a new power of temperature
fluctuations generated by the weak lensing effect on arcminute scales
below the Silk damping scale can be directly used to reconstruct the
projected dark matter field, provided that we have data with
arcminute-scale spatial resolution and sensitivity of $\sim 10 \mu {\rm
K-arcmin}$ (\cite{Seljak}; \cite{Zald}; \cite{HuLens01}) and we can
remove other secondary signals such as the SZ effect by taking advantage
of the specific spectral properties or by combining the CMB measurement
with the $X$-ray or the SZ effect itself. Furthermore, if measurements
of the CMB polarization become possible, combinations of the lensing
effects on the temperature fluctuation and the $E$- and $B$-type
polarization fields can improve the reconstruction (\cite{ZS98};
\cite{HuLens01b}).  This is because the lensing effect induces the
$B$-type polarization on the observed sky from the non-linear coupling
to the primary $E$-mode, evin if the primary CMB contains the $E$-mode
only as suggested by starndard inflationary scenarios, and there is
negligible cosmological contamination of the polarization field in the
arcminute regime unlike the temperature fluctuations (\cite{Hu00a}).

The projected dark matter field is
called the convergence field, $\kappa$, and can be expressed in the form
of a weighted projection of the three-dimensional density fluctuation
field along the line of sight;
\begin{equation}
\kappa\equiv \int^{\eta_0}_{\eta_\ast}\!\!
d\eta W(\eta_\ast,\eta)\delta(\bm{x},\eta),
\label{eqn:conv}
\end{equation}
where $W(\eta_\ast,\eta)$ is the lensing weight function defined by
\begin{equation}
W(\eta_\ast,\eta)\equiv \frac{3}{2}\Omega_{\rm m0}H_0^2
\frac{r(\eta-\eta_\ast)}{r(\eta_\ast)}r(\eta)a^{-1}. 
\label{eqn:lensweight} 
\end{equation}
On the other hand, the lensing distortion effect on distant galaxy
images can be similarly used to measure the relatively low redshift dark
matter distribution such as $z\simlt 2$, depending on typical redshift
of source galaxies (\cite{Bartel}). In this case, the lower limit in the
integration (\ref{eqn:conv}) becomes the conformal time corresponding to
the redshift of source galaxies $z_s$.  However, we should bear in mind
that the lensing effect on galaxies is relatively irrelevant for the
purpose of this paper, because this method usually restricts the
convergence field to low redshifts (e.g. see \cite{Ludvic}), while the
reionization signal in the CMB map is imprinted at moderate and high
redshifts.  The weak lensing effects on CMB thus has a great advantage
of probing the dark matter distribution up to high redshift, which
cannot be attained by any other means (see also \cite{TKF}; \cite{TF};
\cite{Takada01}). In this paper, we assume that the convergence field
$\kappa(\bm{\gamma})$ with respect to direction $\bm{\gamma}$ is {\it a
prior} reconstructed from measurements of the lensing effects on CMB.

\subsection{Doppler Effect of $O(v)$ and $O(v\delta_e)$}

From equation (\ref{eqn:fullCMB2nd}), the secondary temperature
fluctuation field induced by the $O(v)$ Doppler effect can be
written as
\begin{equation}
\Delta(\bm{\gamma})=\int^{\eta_0}_{\eta_{\rm ion}}\!\!d\eta g(\eta)
X_e(1+\delta_e)\bm{\gamma}\cdot\bm{v}. 
\label{eqn:1stDp}
\end{equation}
The pure linear order Doppler effect, $\Delta\sim O(v)$, is suppressed
on small scales because the gravitational instability generates
irrotational flow, leading to cancellations between positive and
negative Doppler shifts along the line of sight (\cite{Kaiser84}).  For
this reason, leading contribution to the secondary temperature
fluctuations due to the Doppler effect comes from second order
perturbations that do not suffer from the cancellation like the OV
effect of $O(v\delta_e)$ (\cite{OV}; \cite{Vish87}; \cite{Efstathiou};
\cite{HuWhite}; \cite{Jaffe}) and the patchy ionization effect of
$O(v\delta_X)$ (\cite{Aghanim}; \cite{Knox}; \cite{GH}; \cite{Benson}),
where $\delta_X$ is the spatial inhomogeneities of ionization fraction
$X_e$ as discussed below equation (\ref{eqn:elenum}).

From equations (\ref{eqn:conv}) and (\ref{eqn:1stDp}), a
cross-correlation between the convergence field and the secondary
temperature fluctuation field due to the Doppler effects of $O(v)$ and
$O(v\delta_e)$ depend on the moments $\skaco{\delta v}$ and
$\skaco{\delta v\delta_e}$ of perturbations, respectively, where the
latter moment seems to produce some contributions in the quasi nonlinear
regime as discussed in \S \ref{LSS}.  However, it is clear that, since
those moments has linear dependence on the bulk velocity field $\bm{v}$,
the cross-correlation is suppressed on relevant small angular scales
because of the cancellations between positive and negative contributions
from the bulk velocity field for the ensemble average.  Note that the
two-point correlation function of the OV effect or equivalently its
angular power spectrum arises from the quadratic contribution of the
velocity field as $O(v^2\delta^2)$, whereas it does not suffer from the
cancellation (\cite{OV}; \cite{Vish87}; \cite{Efstathiou}).

\subsection{Quadratic Doppler Effect of $O(v^2)$}
We next consider secondary temperature fluctuations induced by the
quadratic Doppler effect of the bulk velocity. In the case of assuming
the homogeneous ionization history, the secondary fluctuations are
divided into two contributions; the Doppler effect of $O(v^2)$ and the
density-modulated effect of $O(v^2\delta_e)$. As will be shown later,
the effect of $O(v^2\delta_e)$ generally produces dominant contribution
on small angular scales $l\simgt 1000$, where the nonlinear structures
with $\delta\simgt 1$ are important.  From equation
(\ref{eqn:fullCMB2nd}), the secondary fluctuation field of $O(v^2)$ is
given by
\begin{equation}
\Delta^{v^2}\!\!(\bm{\gamma})=\int^{\eta_0}_{\eta_{\rm ion}}\!\!d\eta
g(\eta)X_e \left[-v^2+7(\bm{\gamma}\cdot{\bm{v}})^2\right].
\label{eqn:2Dpaniso}
\end{equation}
The bulk flow in adiabatic CDM models arises mainly from the linear
regime. In the $\Lambda$CDM model, half the contribution to
$\skaco{v^2}$ comes from scales $k\simlt 0.07 h{\rm Mpc}^{-1}$, and the
density fluctuations go nonlinear at $k\simgt 0.15 h{\rm Mpc}^{-1}$.
This is the reason that the secondary CMB anisotropies by the Doppler
effects of $O(v^2)$ and $O(v^2\delta_e)$ are the order of magnitude
smaller than the anisotropies by the OV effect of $O(v\delta_e)$ on
relevant angular scales.  However, from the fact that there is no
cross-correlation between the convergence field and the OV effect, the
quadratic Doppler effects can provide primary contribution to the
cross-correlation between the convergence field and the secondary CMB
temperature fluctuations induced by the ionized medium.

Since we are interested in the cross-correlation on small angular scales
such as $l\simgt 100$, it is useful to employ the small angle
approximation originally developed by Bond \& Efstathiou (1987).  In
this case, the unit vector $\bm{\gamma}$ in equation
(\ref{eqn:2Dpaniso}), which denotes the direction of line of sight, can
be considered to have component as $\bm{\gamma}\approx
(\theta_x,\theta_y,\sqrt{1-\theta^2})$ for $|\theta_x|, |\theta_y|\ll1$
in the small region around the North Pole on the celestial sphere. It is
then convenient to introduce the two-dimensional vector $\bm{\theta}$,
which has $\bm{\theta}=(\theta_x,\theta_y,0)$. Within the context of
the small angle approximation, the cross-correlation function between
the convergence field $\kappa$ and the temperature fluctuation field
$\Delta^{v^2}$ separated by the angle of $\theta$ is defined by
\begin{equation}
C^{\kappa v^2}(\theta)\equiv \skaco{\kappa(\bm{\gamma}_a)
\Delta^{v^2}\!\!(\bm{\gamma}_b)}_{\bm{\gamma}_a\cdot\bm{\gamma}_b=\cos\theta}
\approx \skaco{\kappa(\bm{\theta}_a)
\Delta^{v^2}\!\!(\bm{\theta}_b)}_{|\bm{\theta}_a-\bm{\theta}_b|\approx 
\theta},
\label{eqn:twoGL2Dp}
\end{equation}
where $\bm{\theta}_a$ and $\bm{\theta}_b$ are the two-dimensional
vectors corresponding to unit vectors $\bm{\gamma}_a$ and
$\bm{\gamma}_b$ for $\kappa$ and $\Delta^{v^2}$,
respectively. Furthermore, if using the two-dimensional Fourier
transformation as, for example,
$\kappa(\bm{\theta})=\int\!d^2\bm{l}/(2\pi)^2\kappa_{\bm{l}}
e^{i\bm{l}\cdot\bm{\theta}}$, the cross-correlation function can be
expressed in terms of its angular power spectrum by the Hankel transform
of zeroth order as
\begin{equation}
C^{\kappa v^2}(\theta)=\int\!\!\frac{ldl}
{2\pi}C_l^{\kappa v^2}J_0(l\theta),  
\label{eqn:defangcl}
\end{equation}
where $J_0(x)$ is the zeroth order Bessel function. The angular power
spectrum $C_l^{\kappa v^2}$ can be also expressed in terms of the
two-dimensional Fourier components of $\kappa$ and $\Delta^{v^2}$ as 
\begin{equation}
\skaco{\kappa_{\bm{l}}\Delta^{v^2}_{\bm{l}'}}\equiv C_l^{\kappa v^2}(2\pi)^2
\delta^2(\bm{l}+\bm{l}'). 
\end{equation}

From equations (\ref{eqn:conv}) and (\ref{eqn:2Dpaniso}), the
cross-correlation function (\ref{eqn:twoGL2Dp}) can be further
calculated as
\begin{eqnarray}
C^{\kappa v^2}\!\!(\theta)
&=&\int^{\eta_0}_{\eta_\ast}\!\!d\eta\int^{\eta_{0}}_{\eta_{\rm ion}}
\!\!d\eta^{\prime} W(\eta,\eta_\ast)
g(\eta^{\prime})\kaco{\delta(\eta,\bm{\theta}_a)X_eq_{e}(\eta',\bm{\theta}_b)}
\nonumber \\
&\approx&\skaco{X_e}\int \!\!d\eta\int\!\!d\eta^{\prime} W(\eta,\eta_\ast)
g(\eta^{\prime})\kaco{\delta(\eta,\bm{\theta}_a)q_{e}(\eta',\bm{\theta}_b)},
\label{eqn:calbi}
\end{eqnarray}
where $q_e(\bm{\theta}_b)\equiv -v^2+7(\bm{\gamma}_b\cdot\bm{v})^2$ with
$\gamma_{b}\approx (\theta_{bx},\theta_{by},1)$, and the functions $g$
and $W$ are given by equations (\ref{eqn:visib}) and
(\ref{eqn:lensweight}), respectively. In the second line on the r.h.s we
have replaced the ionization fraction of $X_e$, which generally depends
on conformal time $\eta$, with the average fraction, $\skaco{X_e}$,
between $\eta_{\rm ion}\le\eta\le\eta_0$ assuming that $X_{\rm
e}(\eta)$ is a slowly varying function with respect to conformal time
after a given reionized epoch $z_{\rm ion}$.  From equation
(\ref{eqn:pert}), the ensemble average
$\skaco{\delta(\bm{\theta}_a,\eta)q_e}$ can be rewritten in terms of
third order moments of perturbations as
\begin{eqnarray}
\skaco{\delta(\bm{\theta}_a)q_e(\bm{\theta}_b)}&\approx&
2\skaco{\delta^{(1)}(\bm{\theta}_a)[-v^{(1)}_iv^{(2)}_i
+7v^{(1)}_zv^{(2)}_z](\bm{\theta}_b)}\nonumber\\
&&+\skaco{\delta^{(2)}(\bm{\theta}_a)
[-v^{(1)}_iv^{(1)}_i+7v_z^{(1)}v^{(1)}_z](\bm{\theta}_b)}+O((\delta^{(1)})^6). 
\end{eqnarray}
Here we have used $\skaco{\delta^{(1)}v^{(1)}v^{(1)}}=0$ and, from the
fact that the propagation direction of CMB photons is almost parallel to
$\hat{z}$-direction in our setting, we have used
$\bm{\gamma}_b\cdot\bm{v}\approx v_z$, where $v_z$ denotes the $z$
component of bulk velocity field.  Therefore, using equations
(\ref{eqn:pecvel}) and (\ref{eqn:2pert}), we can obtain
\begin{eqnarray}
\skaco{\delta(\bm{\theta}_a)q_e(\bm{\theta}_b)}&=&D(\eta)[D\dot{D}^2](\eta')
\int\!\!\frac{d^3\bm{k}}{(2\pi)^3}\int\!\!\frac{d^3\bm{k}'}{(2\pi)^3}
\frac{7k'_z(k_z+k'_z)-k^{\prime 2}-\bm{k}\cdot\bm{k}'}
{k^{\prime 2}|\bm{k}+\bm{k}'|^2}\nonumber\\
&&\times \left[4G(-\bm{k},\bm{k}+\bm{k}')+F(-\bm{k}',\bm{k}+\bm{k}')
\right]P(k)P(|\bm{k}+\bm{k}'|)e^{i\bm{k}_\perp\cdot
(r\bm{\theta}_a-r'\bm{\theta}_b)-ik_z(\eta-\eta')}, 
\label{eqn:GLv2}
\end{eqnarray}
where $\bm{k}_{\perp}$ are $(x,y)$-components of $\bm{k}$ that are
perpendicular to the line of sight, and $r$ and $r'$ denote the comoving
angular distances from the present time to $\eta$ and $\eta'$,
respectively. One can readily see that the integration of the term
$e^{-ik_z(\eta-\eta')}$ over $k_z$ on the r.h.s of equation
(\ref{eqn:GLv2}) leads to the Dirac delta function $\delta(\eta-\eta')$
as long as we are concerned with wavelength modes much smaller than the
comoving Hubble distance characterized by $\eta$. This leads to the
consequence that dominant contribution to the cross-correlation function
arises from structures of dark matter and ionized medium that are
distributed at the same redshift and are separated by $\theta$ on the
observed sky.  This is the so-called the Fourier-space analogue of
Limber's equation (\cite{Kaiser92}; \cite{Jaffe}).  It is known that
this flat sky approximation causes errors no greater than $O(0.1\%)$
at $l\simgt 1000$ (\cite{Hu00b}; \cite{Jaffe}).  Under this
approximation, the cross-correlation function $C^{\kappa
v^2}\!\!(\theta)$ can be calculated as
\begin{equation}
C^{\kappa v^2}\!\!(\theta)\approx \skaco{X_e}\int^{\eta_0}_{\eta_{\rm ion}}\!\!
d\eta g(\eta)W(\eta,\eta_\ast) D^2\dot{D}^2
\int\!\!\frac{k_\perp dk_{\perp}}{2\pi}S(k_\perp)
J_0(k_\perp r\theta),
\label{eqn:twobisp}
\end{equation}
with
\begin{eqnarray}
S(k)&\equiv&2\int^{\infty}_0\!\!dk'\int^{2\pi}_0\!\!d\phi\int^1_0\!\!d\mu
\frac{k^{\prime 2}(7\mu^2-1)-k'k\cos\phi\sqrt{1-\mu^2}}
{k^2+k^{\prime 2}+2kk'\cos\phi\sqrt{1-\mu^2}}\nonumber\\
&&\times\left[4G(k,k',\phi,\mu)+F(k,k',\phi,\mu)\right]P(k')
P(\sqrt{k^2+k^{\prime 2}+2kk'\cos\phi\sqrt{1-\mu^2}}),
\end{eqnarray}
where we have assumed that $\bm{k}_\perp$ is along the $x$-coordinate
from the statistical isotropy and then
$\bm{k}_{\perp}\cdot\bm{k}'=k_\perp k'\cos\phi\cos\theta$.

Hence, from equations (\ref{eqn:defangcl}) and (\ref{eqn:twobisp}) we
can finally derive the angular power spectrum of the cross-correlation
between the convergence field and the secondary temperature fluctuation
field by the quadratic Doppler effect of $O(v^2)$:
\begin{equation}
C_l^{\kappa v^2}=\skaco{X_e}
\int^{\eta_0}_{\eta_\ast}\!\!d\eta g(\eta) W(\eta,\eta_\ast)
r^{-2}(\eta)\dot{D}^2D^2 S\!\!\left(k=\frac{l}{r}\right).
\label{eqn:clbis}
\end{equation}
This is the first main result of this paper. It should be noted that the
integrand quantity in $C^{\kappa v^2}_l$ depends on the cosmological
parameters and on the dark matter power spectrum. Namely, it is not
affected by the unknown biasing relation between the dark matter and
ionized medium distributions.  In this sense, we can accurately predict
the magnitude of the integrand quantity once we fix the dark matter
power spectrum such as the CDM model and the cosmological parameters,
which will be well constrained in the near future by other observations
such as measurements of the CMB anisotropies, the weak lensing survey
and so on.  We therefore expect that measurements of $C_l^{\kappa v^2}$
can set precise constraints on the average ionization fraction
$\skaco{X_e}$ between the reionization epoch and present.

\subsection{Density-Modulated Quadratic Doppler Effect of $O(v^2\delta_e)$}
Next, we consider the secondary temperature fluctuation field induced by the
density-modulated quadratic Doppler effect of $\Delta\sim O(\delta_e
v^2)$. From equation (\ref{eqn:fullCMB2nd}) we have
\begin{equation}
\Delta^{\delta v^2}\!\!(\bm{\theta})=\int\!\!d\eta g(\eta)X_e \delta_e\left[
-v^2+7(\bm{\gamma}\cdot\bm{v})^2\right].
\end{equation}
In the following, to relate $\delta_e$ to the dark matter fluctuation
field $\delta$, we introduce the linear biasing parameter $b$ expressed as
$\delta_e(\bm{x},\eta)\equiv b\delta(\bm{x},\eta)$ for simplicity. 
Note that $b$ generally depends on time and spatial position. 

It is relatively straightforward to calculate the cross-correlation
function between the convergence field and the temperature fluctuations
field $\Delta^{\delta v^2}$:
\begin{eqnarray}
C^{\kappa ,\delta v^2}\!(\theta)&\equiv& \skaco{\kappa(\bm{\theta}_a)
\Delta^{\delta v^2}\!\!(\bm{\theta}_b)}=\int\!\!d\eta \int\!\!d\eta'
W(\eta,\eta_\ast)g(\eta')
\kaco{X_e b\delta(\eta,\bm{\theta}_a)[\delta\left\{-v^2+7(\bm{\gamma}_b\cdot
\bm{v})^2\right\}](\eta',\bm{\theta}_b)}\nonumber\\
&\approx& \skaco{X_e b}\int\!\!d\eta \int\!\!d\eta'
W(\eta,\eta_\ast)g(\eta')
\kaco{\delta(\eta,\bm{\theta}_a)[\delta\left\{-v^2+7(\bm{\gamma}_b\cdot
\bm{v})^2\right\}](\eta',\bm{\theta}_b)} \nonumber\\
&=&\skaco{X_eb}\int\!\!d\eta \int\!\!d\eta'
W(\eta,\eta_\ast)g(\eta')
\skaco{\delta\delta}\skaco{-v^2+7(\bm{\gamma}_b\cdot\bm{v})^2}\nonumber\\
&=&\skaco{X_eb}\int\!\!d\eta \int\!\!d\eta'
W(\eta,\eta_\ast)g(\eta')
\skaco{\delta(\eta,\bm{\theta}_a)\delta(\eta',\bm{\theta}_b)}
\frac{4}{3}\skaco{v^2},
\end{eqnarray}
where $\skaco{X_e b}$ is the average quantity of $X_e b$ between the
reionization epoch and present. After the similar procedure as used in the
derivation of equation (\ref{eqn:clbis}), we can obtain the angular
power spectrum
\begin{equation}
C_l^{\kappa,\delta v^2}\approx 4\skaco{X_e b} 
\int^{\eta_0}_{\eta_{\rm ion}} 
\!\!d\eta W(\eta,\eta_\ast)g(\eta)r^{-2}(\eta)D^2 P\!
\left(k=\frac{l}{r}\right)v^2_{\rm rms}(\eta),
\label{eqn:cldens}
\end{equation}
where $v_{\rm rms}$ is the rms of the present-day bulk velocity
field and can be expressed in terms of the linear density power 
spectrum as
\begin{equation}
v^2_{\rm rms}(\eta)=\frac{\dot{D}^2(\eta)}{3}
\int\!\!\frac{dk'}{2\pi^2}P(k'). 
\label{eqn:rmsvel}
\end{equation}
Here the factor $1/3$ is due to the three-dimensional freedom of bulk
velocity field such that $v_{\rm rms}$ represents the rms of
one-dimensional motion.  The most important result of equation
(\ref{eqn:cldens}) is that the angular power spectrum
$C_l^{\kappa,\delta v^2}$ is proportional to not quadratic but linear
term of $\skaco{X_e b}$ for all $l$ and therefore could be both positive
and negative, depending on the sign of $\skaco{X_e b}$ or more
specifically on the sing of $\skaco{b}$, since the integrand quantity in
equation (\ref{eqn:cldens}) is always positive.  The quantity
$\skaco{X_e b}$ should be a crucial clue to understanding of how the
ionized medium distribution is related to the dark matter distribution
in the large-scale structure, which still remains large uncertainties
both from theoretical and observational aspects.  
The measurements of $C_l^{\kappa,\delta v^2}$
can thus be a new direct probe of revealing the projected biasing
relation.  It is worth noting that the angular power spectrum of the OV
effect depends on the quadratic term $O(X_e^2 b^2)$ and is always
positive irrespective of the sing of $\skaco{X_eb}$.

The simple assumption that the ionized medium traces the dark matter
distribution allows us to place the upper limit on $C_l^{\kappa,\delta
v^2}$ using the nonlinear dark matter power spectrum, which has been
well studied with N-body simulations for the full range of adiabatic CDM
models.  Equation (\ref{eqn:cldens}) can be then rewritten as
\begin{equation}
C_l^{\kappa,\delta v^2}=4\skaco{X_eb}
\int^{\eta_0}_{\eta_{\rm ion}}
\!\!d\eta W(\eta,\eta_\ast)g(\eta)r^{-2}
P_{\rm NL}\left(k=\frac{l}{r},\eta\right)v^2_{\rm rms}(\eta),
\label{eqn:cldensNL}
\end{equation}
where $P_{\rm NL}(k,\eta)$ is the power spectrum of nonlinear density
fluctuations for which we will employ the fitting formula developed by
Peacock \& Dodds (1996).  Note that we here consider a possible case
that electrons in reionized medium with non-linear density fluctuations
are distributed along the line of sight moving according to a
large-scale coherent bulk velocity field, as suggested by the CDM power
spectrum. This is in fact an essential mechanism that the
Ostriker-Vishniac effect can avoid cancellations between blue- and red
shifts along the line of sight from the coupling of the density and
velocity fields.  Based on this consideration, in Figure \ref{fig:cl_NL}
we use the non-linear power spectrum only for the calculation of
$C_l^{\kappa,\delta v^2}$ and use the second-order perturbation theory
for that of $C_l^{\kappa v^2}$, because the latter arises from the
ensemble average of the density fluctuation field and the bulk velocity
filed on the mildly non-linear regime.

\section{RESULTS}
\label{result}

We present numerical results for the angular power spectra given by
equations (\ref{eqn:clbis}) and (\ref{eqn:cldens}) for the $\Lambda$CDM
model.  We adopt a set of cosmological parameters of $\Omega_{\rm
m0}=0.3$, $\Omega_{\lambda 0}=0.7$, $\Omega_{\rm b0}=0.05$, $h=0.7$, and
$\sigma_8=1$, which is consistent with the recent CMB experiments
(\cite{Boom2}) and the cluster abundance (\cite{Eke}; \cite{KS}).  We
simply consider the homogeneous hydrogen ionization model after a given
reionization epoch in the range of $5<z_{\rm ion}<30$. In this case we
will consider $X_e={\rm const.}$ with respect to time during $0\le z\le
z_{\rm ion}$ and equations (\ref{eqn:opt}) and (\ref{eqn:visib}) can be
used for calculations of the optical depth and visibility function,
respectively. Although more realistic model is a patchy or inhomogeneous
reionization (e.g., \cite{Benson}), our model allows us to compute the
magnitude of effects in the simplest way.

We first estimate the order of magnitude of secondary temperature
fluctuations by the quadratic Doppler effect.  Using the $\Lambda$CDM
model above, the present-day rms of bulk velocity field can be estimated
from equation (\ref{eqn:rmsvel}) as $v_{\rm rms}(t_0)\approx 1.19\times
10^{-3}$ in the unit of $c=1$.  Equation (\ref{eqn:fullCMB2nd}) tells us
that the magnitude of temperature fluctuations induced by the quadratic
Doppler effect can be roughly estimated as $\Delta^{v^2} \sim \tau
v_{\rm rms}^2 \sim 10^{-9}-10^{-8}$ for $z_{\rm ion}=10$, since we have
$\tau \sim 10^{-3}-10^{-2}$ for the model considered in this paper.
Similarly, the magnitude of temperature fluctuations induced by the OV
effect can be estimated as $\Delta^{\rm OV}\sim \tau v_{\rm rms}\delta
\sim 10^{-6}$, which is consistent with the numerical results done by Hu
\& White (1996) (see also \cite{Efstathiou} and \cite{Jaffe}). We thus
conclude $\Delta^{v^2}\ll \Delta^{\rm OV}$ in the observed CMB map
itself, and the magnitude of $\Delta^{v^2}$ is much below sensitivities
of the satellite missions {\em MAP} and {\em Planck Surveyor}.

\begin{figure}[t]
 \begin{center}
     \leavevmode\epsfxsize=12cm \epsfbox{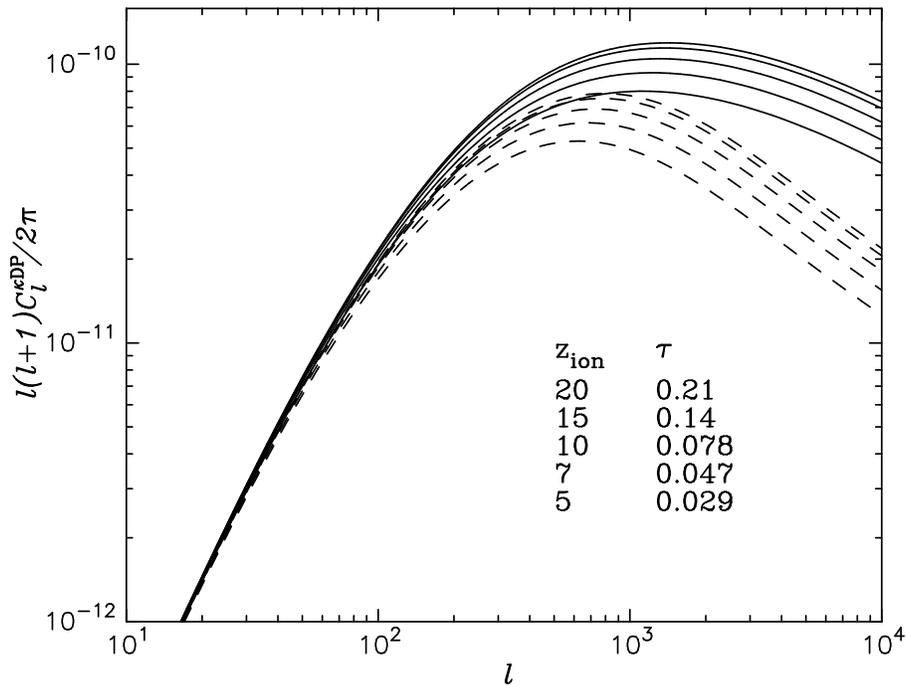}
\caption{ Angular power spectra of the cross-correlation between the
weak lensing field $\kappa$ and the secondary CMB fluctuations field
$\Delta$ induced by the quadratic Doppler effects of ionized medium.
These spectra are calculated under the simple assumptions that the gas
traces the dark matter and the full hydrogen ionization after a given
reionization epoch; $X_e=1$ and $b=1$ for the ionization fraction and
the biasing parameter.  The solid and dashed lines show contributions
from the pure quadratic Doppler effect of $O(v^2)$ and the
density-modulated effect of $O(v^2\delta_e)$ given by equations
(\ref{eqn:clbis}) and (\ref{eqn:cldens}), respectively. We here
considered five cases of the reionization epoch as $z_{\rm
ion}=5,7,10,15,20$.  \label{fig:cl}}
 \end{center}
\end{figure}

Figure \ref{fig:cl} shows the angular power spectra of the
cross-correlations between the weak lensing field and the secondary
temperature fluctuations induced by the quadratic Doppler effect of
$O(v^2)$ (dashed lines) and the density-modulated effect of $O(\delta_e
v^2)$ (solid), which are computed using equations (\ref{eqn:clbis}) and
(\ref{eqn:cldens}), respectively. We here assumed $b=1$ for the unknown
biasing parameter between the dark matter and electron distributions in
the large-scale structure.  Both the magnitudes of the Doppler effects
of $O(v^2)$ and $O(\delta_e v^2)$ are comparable on $l\simlt 10^2$ in
the case of no biasing $b=1$. This is because leading contribution to
the cross-correlation between the weak lensing field of $O(\delta)$ and
the Doppler effect of $O(v^2)$ comes from the fourth-order moments of
linear perturbations like $O(\delta^{(1)2}v^{(1)2})$ resulting from the
vanishing of the third-order moments $\skaco{\delta^{(1)}v^{(1)2}}=0$,
while the cross-correlation from the density-modulated Doppler effect of
$O(\delta_e v^2)$ also arises from the moments of
$\skaco{\delta^{(1)}v^{(1)2}}$.  It is clear that the contribution from
the density-modulated effect of $O(v^2\delta_e)$ dominates the effect of
$O(v^2)$ on angular scales of $l\simgt 100$. The total magnitudes of the
angular power spectra peak $C^{\kappa {\rm DP}}_l l^2/2\pi\sim 10^{-10}$
around $l\approx 1500$ for the $\Lambda$CDM model and for the five cases
of reionization epoch $z_{\rm ion}=5,7,10,15,20$. The peak location in
$l$-space results from the shape of CDM power spectrum.  On the other
hand, Cooray (2000) investigated the angular power spectrum of the
cross-correlation between the weak lensing field and the thermal SZ
temperature fluctuation field based on the halo approach of dark matter
clustering (\cite{KK}) and the simple model of intracluster gas
distribution.  The magnitude of the angular power spectrum from the SZ
effect amounts to $C_l^{\kappa {\rm SZ}}l^2\sim 10^{-8}$ at $l\approx
2000$ for the similar models as considered in this paper.  The
temperature fluctuation induced by the quadratic Doppler effect is over
an order of magnitude smaller than that by the SZ effect because of
$\Delta^{v^2}\sim 10^{-9}-10^{-8}$ and $\Delta^{\rm SZ}\sim
10^{-6}$. From this fact, it seems unlikely that the magnitude of the
cross-correlation between $\kappa$ and the quadratic Doppler effect is
smaller only by $\sim 10^{-2}$ than that from the SZ effect.  This is
because the cross-correlation with the Doppler effect comes from
wide-range redshift structures such as $0<z<z_{\rm ion}$, while the SZ
effect has dominant contributions from low redshift massive dark halos
such as clusters of galaxies with $M\simgt 10^{14}M_\odot$ at $z\simlt
0.5$ (\cite{Cooray}). 
These results therefore imply that, if the weak lensing field is {\it a
priori} reconstructed by measurements of lensing effects on the CMB, the
cross-correlation could improve the S/N of the quadratic Doppler effect,
even though the angular power spectrum itself of the Doppler effect has a
small power of $C_l^{\rm DP}l^2\simlt 10^{-17}$ only. This is an
analogous statement to the cross-correlation between the primary CMB
temperature fluctuation and polarization fields (e.g.,
\cite{HuWhite97}).

\begin{figure}[t]
 \begin{center}
     \leavevmode\epsfxsize=12cm \epsfbox{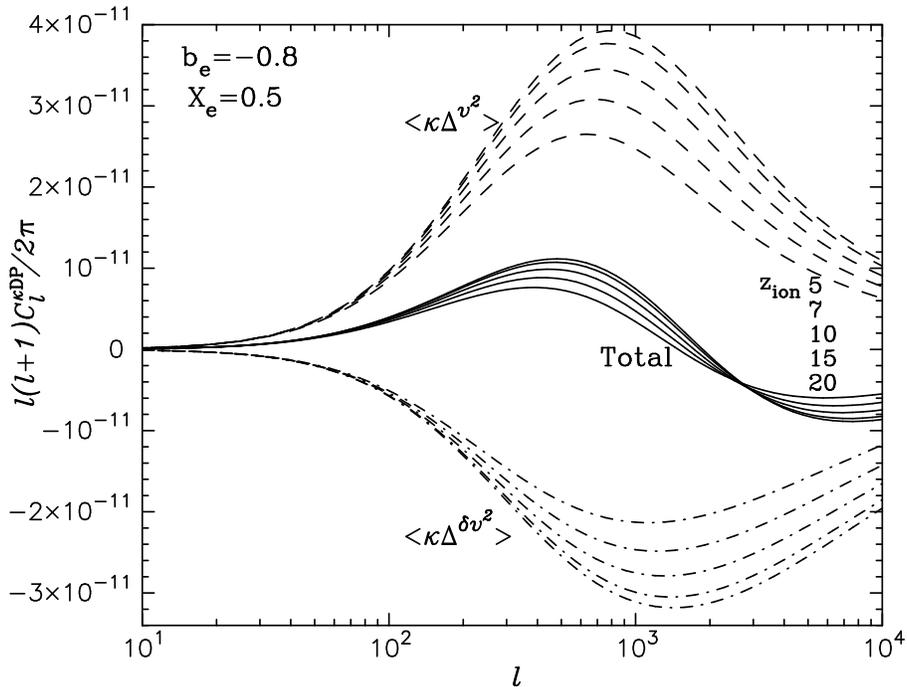}
\caption{ A toy model of cross-correlation angular power spectrum for
the simplest model of antibiasing relation between the dark matter and
electron distributions given by $b=-0.8$. We here considered the
reionization epochs of $z_{\rm ion}=5,7,10,15,20$ and assumed $X_e=0.5$
for the ionization fraction after the respective reionization epoch.
The dashed and dotted-dashed lines show the angular power spectra of
cross-correlation between the weak lensing field and the temperature
fluctuation field induced by the Doppler effects of $O(v^2)$ and
$O(v^2\delta_e)$, respectively, as shown in Figure \ref{fig:cl}.  Solid
lines show the total angular power spectra that we would actually
observe, and the spectra become negative on $l\simgt 2000$ for this
model.  \label{fig:antibias}}
 \end{center}
\end{figure}

To explicitly illustrate how the cross-correlation considered could be
useful to reveal the biasing relation between the dark matter and
electron distributions, we here employ a toy model of the antibiasing
relation, motivated by the reionization scenario of low-density region
prior to the high-density region as suggested by MHR. The homogeneous
full hydrogen ionization with $X_e=1$ seems to lead to
$\delta_e=\delta_{b}$ from $n_e=n_{\rm H}$, which seems inconsistent
with the antibiasing relation, and thus we here assume the partly
homogeneous ionization fraction of $X_e=0.5$ after $z_{\rm ion}$.
Although the antibiasing process depends on how reionization proceeds in
the large-scale structure in detail, we consider the simplest model of
linear and constant antibiasing relation given by $\delta_e=b\delta$
with $b=-0.8$.  It should be then noted that the condition of
$\delta_e\ge -1$ leads to $\delta<1/|b|$. Accordingly, we have taken
into account the range of $\delta$ for the calculation of the ensemble
average of $\skaco{\delta_e\delta}$ used in equation (\ref{eqn:cldens}).
Figure \ref{fig:antibias} shows the result of this toy model, and one
can readily see that we should observe the negative total angular power
spectra (solid lines) of the cross-correlation between the weak lensing
field and the secondary temperature fluctuation field induced by the
Doppler effect on $l\simgt 2000$.  Note that this critical scale of $l$
depends on the value of the antibiasing parameter $b$, because the
cancellation between the spectra from the quadratic Doppler effect of
$O(v^2)$ and the density-modulated effect of $O(v^2\delta_e)$ also
depends on $b$.  As expected, the cross-correlation function could thus
be a direct probe of the antibiasing relation, which cannot be attained
by measurements of the angular power spectrum of the $O(V)$ Doppler
effect such as the OV effect because of the quadratic dependence on the
biasing relation.

\begin{figure}[t]
 \begin{center}
     \leavevmode\epsfxsize=12cm \epsfbox{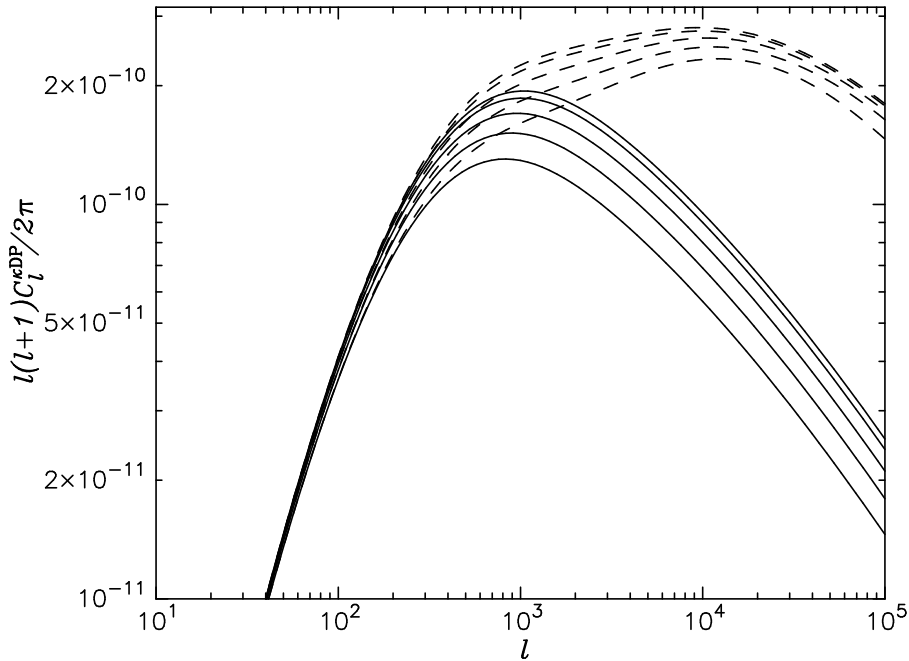}
\caption{ Maximal nonlinear enhancement of the angular power spectra of
the cross-correlations for the same models ($X_e=b=1$) as shown in
Figure \ref{fig:cl}. Solid and dashed lines demonstrate the total
angular power spectra taking into using the linear and nonlinear dark
matter power spectrum, respectively.  These power spectra includes both
contributions from the cross-correlations of the weak lensing field with
the pure quadratic Doppler effect of $O(v^2)$ and the
density-modulated effect of $O(v^2\delta_e)$ as shown in Figure
\ref{fig:cl}. We used equation (\ref{eqn:cldensNL}) for the calculation
of the nonlinear density-modulated spectrum.  Nonlinear effect could be 
important at $l\simgt 1000$.  \label{fig:cl_NL}}
 \end{center}
\end{figure}

Figure \ref{fig:cl_NL} shows the maximal estimate of the angular power
spectrum of the cross-correlation using the nonlinear dark matter power
spectrum under the simplest assumption that the gas traces dark matter
distribution in the large-scale structure. We here again considered the
simplest model of $X_e=1$ and $b=1$ as in Figure \ref{fig:cl}. The
nonlinear enhancement could be important on small angular scales of
$l\simgt 10^3$. However, it is not still clear if the electron
distribution traces dark matter distribution in the nonlinear regime,
because physical processes of the gas pressure and radiative dynamics
should also play an important role in such nonlinear regions (e.g.,
\cite{Gnedin00}).  For this reason, the curves in Figure \ref{fig:cl_NL}
provide us with upper limits on a realistic angular power spectrum of the
cross-correlation.

%

\section{DISCUSSION AND CONCLUSIONS}
\label{Conc}

In this paper, we have investigated the cross-correlation between the
weak lensing field and the secondary CMB temperature fluctuations
induced by the Doppler effect of the ionized medium in the reionized
epoch of the universe.  The main result of this paper is to develop a
formalism for calculating the cross-correlation function and its angular
power spectrum based on the small angle approximation. 
The leading order to the cross-correlation comes from the secondary CMB
anisotropies due to the quadratic Doppler effects of the bulk motions,
since the cross-correlation with the $O(v)$ Doppler effect such as the
OV effect depends linearly on the bulk velocity field and thus suffers
from cancellations between positive and negative contributions from the
bulk velocity field for the ensemble average.  There are two main
contributions to the cross-correlation function: one is contribution
from the pure quadratic Doppler effect of $O(v^2)$ and the other is that
from the density-modulated effect of $O(v^2\delta_e)$, where $\delta_e$
is the electron density fluctuation field.  For the CDM models, the
latter contribution dominates the former on angular scales of $l\simgt
100$ if the ionized medium traces the dark matter distribution.  We
found that the cross-correlation of $O(X_ev^2\delta)$ from the Doppler
effect of $O(v^2)$ is sensitive to the average ionization fraction
$\skaco{X_e}$ between the reionization epoch and present in the sense
that the bulk velocity and the weak lensing field depends only on the
dark matter fluctuations, through the gravitational instability
scenario, and on the cosmological parameters, which will be well
constrained by the future precise data such as the primary CMB
anisotropies and the weak lensing survey.  On the other hand, we showed
that the cross-correlation of $O(X_ev^2\delta_e\delta)$ between the weak
lensing field and the density-modulated quadratic Doppler effect of
$O(v^2\delta_e)$ depends linearly on the biasing relation between the
dark matter and free electron distributions, and therefore its angular
power spectrum could be both positive and negative, depending on the
biasing and antibiasing, respectively. If the low density region of IGM
is ionized prior to the high density region in the reionization history
as suggested by MHR (see also \cite{Gnedin00}), we would measure the
negative angular spectrum of the cross-correlation.  Our results thus
offer a new opportunity to understand the unknown problem how
reionization proceeds in the inhomogeneous intergalactic medium
connected to the dark matter clustering in principle, although the
magnitude of the signal is very small.

Let us briefly comment on other sources of cross-correlation between the
weak lensing field and the observed CMB map. One significant source is
the thermal SZ effect, and it has been indeed shown that the
cross-correlation could have the magnitude of $\skaco{\Delta^{\rm
SZ}\kappa}\sim 10^{-8}$ for the similar CDM model (\cite{Cooray}) as
investigated in this paper.  The SZ effect thus produces much larger
contribution than the cross-correlation with the Doppler effect
estimated as $\skaco{\Delta^{\rm DP}\kappa}\sim 10^{-10}$. However,
since the thermal SZ signal arises mainly from hot plasma in clusters of
galaxies at low redshift $(z\simlt 0.5)$, it is expected that we can
remove this signal from the cross-correlation maps, combined with the
X-ray data or the SZ maps. The other possible sources are the CMB
foregrounds such as synchrotron radiation from extragalactic dust that
may correlate with the dark matter distribution in the large-scale
structure. Isolating the signal shown in this paper observationally from
the foregrounds will be a great challenge by using the different
frequency-dependence (\cite{TegFore}), though this problem is beyond the
scope of this paper.

\section*{Acknowledgments}
M.~T. acknowledges financial support from the Japan Society for
Promotion of Science (JSPS) Research Fellowships.  This research was
supported by Japanese Grant-in-Aid for Science Research Fund of the
Ministry of Education, Science, Sports and Culture Grant Nos. 06107 and
11640235, and Sumitomo Fundation.


\begin{thebibliography}{}
\parskip=0pt \baselineskip=7pt
\bibitem[Aghanim et al. 1996]{Aghanim}
Aghanim, N., D\'esert, F.~X., Puget, J.~L., \& Gispert, R. 1996, 
\aap, 311, 1
\bibitem[Becker et al. 2001]{Becker}
Becker, R.~H., et al. 2001, astro-ph/0108097
\bibitem[Bardeen et al. 1986]{BBKS}
   Bardeen, J. M., Bond., J. R., Kaiser, N., \& Szalay, A. S. 1986,
                              \apj, 304, 15 (BBKS)
\bibitem[Bartelmann, \& Schneider 2001]{Bartel}
Bartelmann, M. \& Schneider, P. 2001, Phys.~Rep. 2001, 340, 291 
\bibitem[Barkana, \& Loeb 2001]{Barkana}
Barkana, R., \& Loeb, A. 2001, Phys.~Rep., in press, astro-ph/0010468
\bibitem[Benson et al. 2001]{Benson}
Benson, A.~J., Nusser, A., Sugiyama, N., \& Lacey, C.~G. 2001, 
\mnras, 320, 1
\bibitem[de Bernardis, et al. 2000]{Boom1}
de Bernardis, et al., P., et al. 2000, \nat, 404, 955 
\bibitem[Bond, \& Efstathiou 1987]{Bond}
Bond, J.~R., \& Efstathiou, G.~P.  1987, \mnras, 226, 655 
\bibitem[Bond et al. 1991]{Bond91}
Bond, J.~R., Efstathiou, G.~P., Lubin, P.~M., \& Meinhold, P.~R. 1991, 
\prl, 66, 2179
\bibitem[Cooray 2000]{Cooray}
Cooray, A. 2000, \prd, 62,  103506
\bibitem[Dedelson, \& Jubas 1995]{Dodelson}
Dodelson, S., \& Jubas, J.~M. 1995, \apj, 439, 503
\bibitem[Efstathiou 1988]{Efstathiou}
Efstathiou, G.~P. 1988, in Large-Scale Motions, edited by V. Rubin and 
S.~J.~Coyne (Princeton Univ. Press)  
\bibitem[Eke, Cole, \& Frenk 1996]{Eke}
 Eke, V., Cole, S., \& Frenk, C. S. 1996, \mnras, 282, 263
\bibitem[Fan, et al. 2000]{Fan}
Fan, X., et al. 2000, \aj, 120, 1167
\bibitem[Gnedin 2000]{Gnedin00}
Gnedin, N.~Y. 2000, \apj, 535, 530
\bibitem[Gnedin, \& Jaffe 2001]{Gnedin01}
Gnedin, N.~Y., \& Jaffe, A.~H. 2001, \apj, 551, 3
\bibitem[Gruzinov \& Hu 1998]{GH}
Gruzinov, A., \& Hu, W. 1998, \apj, 508, 435,
\bibitem[Gunn, \& Peterson 1965]{Gunn}
Gunn, J.~E., \& Peterson, B.~A. 1965, \apj, 142, 1633
\bibitem[Jaffe, \& Kamionkowski 1998]{Jaffe}
Jaffe, A., \& Kamionkowski, M. 1998, \prd, 58, 043001
\bibitem[Hanany et al. 2000]{Maxima}
Hanany, S., et al. 2000, \apj, 545, L5
\bibitem[Hu 2000a]{Hu00a}
Hu, W. 2000, \apj, 529, 12
\bibitem[Hu 2000b]{Hu00b}
Hu, W. 2000, \prd, 62, 043007 
\bibitem[Hu 2001a]{HuLens01}
Hu, W. 2001a, \apj, 557, L79
\bibitem[Hu 2001b]{HuLens01b}
Hu, W. 2001b, astro-ph/0111606
\bibitem[Hu, \& White 1996]{HuWhite}
Hu, W., \& White, M. 1996, \aap, 315, 33
\bibitem[Hu, \& White 1997]{HuWhite97}
Hu, W., \& White, M. 1997, \prd, 56, 596
\bibitem[Hu, Scott, \& Silk 1994]{Hu94}
Hu, W., Scott, D., \& Silk, J. 1994, \prd, 49, 648 (HSS)
\bibitem[Kaiser 1984]{Kaiser84}
Kaiser, N. 1984, \apj, 282, 374
\bibitem[Kaiser 1992]{Kaiser92}
Kaiser, N. 1992, \apj, 388, 272
\bibitem[Kitayama \& Suto 1997]{KS}
         Kitayama, T., \& Suto, Y. 1997, \apj, 490, 557
\bibitem[Knox, Scoccimarro, \& Dodelson 1998]{Knox}
Knox, L., Scoccimarro, R., \& Dodelson, S. 1998, \prl, 81, 2004
\bibitem[Komatsu, \& Kitayama 1999]{KK}
Komatsu, E., \& Kitayama, T. 1999, \apj, 526, L1
\bibitem[Lanzetta, Wolfe, \& Turnashek 1995]{Lanzetta}
Lanzetta, K.~M., Wolfe, A.~M., \& Tunshek D.~A. 1996, \apj, 440, 435
\bibitem[Miralde-Escude, Haehnelt, \& Rees 2000]{Miralde}
Miralde-Escude, J., Haehnelt, M., \& Rees, M. 2000, \apj, 530, 1 (MHR)
\bibitem[Netterfield, et al. 2001]{Boom2}
Netterfield, C.~B., et al. 2001, astro-ph/0104460
\bibitem[Ostriker, \& Vishniac 1986]{OV}
Ostriker, J.~P., \& Vishniac, E.~T. 1986, \apj, 306, L51
\bibitem[Peacock \& Dodds]{PD} 
Peacock, J. A., \& Dodds, S. J. 1996, \mnras, 280, L19
\bibitem[Peebles 1980]{Peebles}
Peebles, P.~J.~E. 1980, The Large Scale Structure of the Universe 
(Princeton Univ. Press)
\bibitem[Peiris, \& Spergel 2000]{Peiris}
Peiris, H.~V., \& Spergel, D.~N. 2000, \apj, 540, 605
\bibitem[Schmalzing, Takada, \& Futamase 2000]{STF}
Schmalzing, J., Takada, M., \& Futamase, T. 2000, \apj, 544, L83
\bibitem[Seljak \& Zaldarriaga 1999]{SZPRL}
 Seljak, U., \& Zaldarriaga, M., 1999, \prl, 82, 2636
\bibitem[Seljak \& Zaldarriaga 2000]{Seljak}
Seljak, U., \& Zaldarriaga, M. 2000, \apj, 538, 57
\bibitem[Sugiyama 1995]{Sugiyama}
 Sugiyama, N. 1995 \apjs, 100, 281
\bibitem[Sugiyama, Silk, \& Vittorio 1993]{SugSilk}
Sugiyama, N., Silk, J., \& Vittorio, N. 1993, \apj,  419, L1
\bibitem[Takada 2001]{Takada01}
Takada, M. 2001, \apj, 558, 29
\bibitem[Takada, \& Futamase 2001]{TF}
Takada, M., \& Futamase, T. 2001, \apj, 546, 620
\bibitem[Takada, Komatsu, \& Futamase 1999]{TKF}
Takada, M., Komtasu, E., \& Futamase, T. 1999, \apj, 533, L83
\bibitem[Tegmark, \& Zaldarriaga 2000]{Teg}
Tegmark, M., \& Zaldarriaga, M. 2000, \prl, 85, 2240
\bibitem[Tegmark, et al. 2000]{TegFore}
Tegmark, M., Eisenstein, D.~J., Hu, W., \& de Oliveira-Costa, A.~O. 
2000, \apj, 530, 133
\bibitem[Vishniac 1987]{Vish87}
Vishniac, T. 1987, \apj, 322, 597
\bibitem[Van Waerbeke, Bernardeau, \& Benabed 2000]{Ludvic00}
Van Waerbek, L., Bernardeau, F., \& Benabed, K. 2000, \apj, 540, 14
\bibitem[Van Waerbeke et al. 2000]{Ludvic}
Van Waerbeke, L. et al. 2000, \aap, 358, 30
\bibitem[Zaldarriaga 2000]{Zald}
Zaldarriaga, M. 2000, \prd, 62,  063510
\bibitem[Zaldarriaga \& Seljak 1998]{ZS98}
Zaldarriaga, M., \& Seljak, U. 1998, \prd, 58, 023003
\bibitem[Zaldarriaga \& Seljak 1999]{ZS99}
 Zaldarriaga, M., \& Seljak, U. 1999, \prd, 59, 123507
\bibitem[Zel'dovich, \& Sunyaev 1969]{SZ}
Zel'dovich, Ya.~B., \& Sunyaev, R.~A. 1969, Astrophys. Space. Sci., 
4, 301
\end{thebibliography}
\end{document}